\documentclass[superscriptaddress,showpacs,twocolumn]{revtex4}
\usepackage{amssymb}
\usepackage[tbtags]{amsmath}
\usepackage{graphicx}
\usepackage{epsfig,graphicx,times}

\begin{document}
\title{An efficient single-step scheme for manipulating quantum information of two trapped ions beyond the Lamb-Dicke limit}
\author{L.F. Wei}
\thanks{lfwei@riken.jp}
\affiliation{Frontier Research System,  The Institute of Physical
and Chemical Research (RIKEN), Wako-shi, Saitama, 351-0198, Japan}
\affiliation{Department of Physics, Shanghai Jiaotong University,
Shanghai 200030, P.R. China }
\author{Franco Nori}
\thanks{fnori@riken.jp}
\affiliation{Frontier Research System, The Institute of Physical
and Chemical Research (RIKEN), Wako-shi, Saitama, 351-0198, Japan}
\affiliation{Center of Theoretical Physics, Physics Department,
Center for the Study of Complex Systems, University of Michigan,
Ann Arbor, Michigan 48109-1120, USA}
\thanks{Permanent address}

\begin{abstract}
Based on the exact conditional quantum dynamics for a two-ion
system, we propose an efficient {\it single-step\/} scheme for
coherently manipulating quantum information of two trapped cold
ions by using a pair of synchronous laser pulses. Neither the
auxiliary atomic level nor the Lamb-Dicke approximation are
needed.


\end{abstract}
\maketitle

\section{Introduction}

The entanglement between different particles has recently become a
focus of activity in quantum physics (see, e.g.,
\cite{Bennett00}), because of experiments on non-local features of
quantum mechanics and the development of quantum information
physics.
Einstein-Podolsky-Rosen (EPR) entangled states with two particles
have been employed not only to test Bell's inequality
\cite{Bell87}, but also to realize quantum cryptography and
quantum teleportation \cite{Bennett00}.
Also, entanglement plays a central quantitative role in quantum
parallelism \cite{Shor94}.
The demonstrations of quantum entanglement to date are usually
based on various probabilistic processes, e.g., the generation of
photon pairs in parametric down conversion \cite{Kwiat95}.
However, it is very difficult to generate the entanglement of
larger numbers of particles, as the probability of randomly
generating the appropriate conditions decreases exponentially with
the number of particles.
Intense activities are now focused on generating an entanglement
of particles in a deterministic way, i.e., to produce a desired
entangled state.
For example, the entanglements of two and four trapped ions have
been produced experimentally \cite{Sackett00}.

As first suggested by Cirac and Zoller \cite{CZ95}, a very
promising scenario for implementing a practical quantum
information processor is the system of laser-cooled trapped ions,
due to its long coherence time \cite{KS01}.
Information in this system is stored in the spin states of an
array of trapped cold ions and manipulated by using laser pulses.
The ions are held apart from one another by their mutual Coulomb
repulsion.
Each ion can be individually addressed by focusing laser beams on
the selected ion.
The collective normal modes of oscillation shared by all of the
ions form the information bus, through which all gate operations
can be performed.
In the past few years, several key features of the proposal in
\cite{CZ95}, including the production of entangled states and the
implementation of quantum controlled operations between a pair of
trapped ions, have already been experimentally demonstrated
\cite{Monroe95,King98,Turchette98,Sackett00}.
Also, several alternative theoretical schemes (see, e.g.,
\cite{Duan01,Jonathan01,wei02,sharma03,Jonathan00,Childs01,MS99})
have been developed for overcoming various difficulties in
realizing a practical ion-trap quantum information processor.


The Lamb-Dicke (LD) approximation is made in {\it almost all} of
these schemes (see, e.g.,
\cite{King98,Sackett00,Jonathan01,Childs01,MS99}), in order to
simplify the treatment of the laser-ion interaction.
This approximation requires that the coupling between the external
and internal degrees of freedom of the ion is very weak, i.e., the
spatial dimension of the motion of the ground state of the trapped
ion should be much smaller than the effective wavelength of the
applied laser field (see, e.g., \cite{Meekhof96}).
Thus, in the LD limit the interaction between the internal states
$|s\rangle=\{|g\rangle,\,|e\rangle\}$ and the external motional
harmonic oscillator states $\{|n\rangle;\,\,n=0,1,2,...\}$ of the
ion can be expanded to the lowest order of the LD parameter
$\eta_L$, then the usual Jaynes-Cummings (JC) or anti-JC-type
model can be derived.
However, the quantum motion of the trapped ions is {\it not\/}
limited to the LD regime \cite{Wineland79,Steane98,Morigi99}.
Inversely, utilizing the laser-ion interaction beyond this limit
could be helpful for reducing the noise in the trap and improving
the cooling rate (see, e.g., \cite{Steane98}). Therefore, it would
be useful to implement the trapped-ion quantum information
processing outside the LD regime.
In fact, several schemes \cite{wei02}\cite{sharma03} have been
proposed to implement the quantum computation with trapped ions
beyond the LD limit by sequently applying a series of pulses.

In this work, we propose an alternative scheme for manipulating
quantum information, e.g. realizing quantum controlled operations
and generating entanglement, of two trapped ions {\it beyond\/}
the LD limit by using a pair of synchronous laser pulses.
The information bus, i.e, the center-of-mass (CM) vibrational
quanta of the ions, for communicating different ions, may be
either in its ground or an arbitrary excitation state.
Neither the auxiliary atomic level nor the Lamb-Dicke
approximation are needed in this work.
The experimental realization of this simple approach is discussed.

\section{Conditional quantum dynamics for two trapped ions
driven by two synchronous classical laser beams beyond the
Lamb-Dicke limit}

The ion trap quantum information processor consists of a string of
ions stored in a very cold linear radio-frequency trap.
The motion of the ions, which are coupled together due to the
Coulomb force between them, is quantum mechanical in nature.
The ions are sufficiently separated apart (see, e.g.
\cite{Steane98,leibfried99}) to be easily addressed by different
laser beams, i.e., each ion can be illuminated individually by a
separate leaser beam.
The communication and logic operations between qubits are usually
performed by exciting or de-exciting quanta of the collective
vibration (i.e., the shared phonon) modes, which act as the
information bus (see, e.g., \cite{CZ95,wei02,Jonathan00}).

%
We consider an array of $N$ two-level cold ions of mass $M$
trapped in a one-dimensional harmonic potential of frequency
$\nu$.
The ions are able to perform small oscillations around their
equilibrium position $z_{i0}\,(i=1,2,...,N)$, due to the repulsive
Coulomb force between them.
Each one of the ions is assumed to be individually addressed by a
separate laser beam.
We consider the case where an arbitrary pair (labeled by $j=1,2$)
of trapped cold ions from the chain of $N$ trapped cold ions are
illuminated independently by two weak classical laser beams.
This is different from the scheme proposed in \cite{Solano00} of
using Raman lasers to drive the ions.
The Hamiltonian corresponding to our situation is
\begin{widetext}
\begin{eqnarray}
\hat{H}(t)&=&\hbar\omega_0\sum_{j=1}^{2}\frac{\hat{\sigma}_{z,j}}{2}+\hbar\,\nu\,(\hat{a}^\dagger\,\hat{a}+\frac{1}{2})
+\sum_{l=1}^{N-1}\hbar\,\nu_l\,(\hat{b}_l^{\dagger}\,\hat{b}_l+\frac{1}{2})\nonumber\\
&+&\frac{\hbar}{2}\sum_{j=1}^{2}\left\{\Omega_j\,\,\hat{\sigma}_{+,j}\exp\{i\,[\eta_j\,(\hat{a}^{\dagger}+\hat{a})
+\sum_{l=1}^{N-1}\eta_{j,\,l}\,(\hat{b}_l^{\dagger}+\hat{b}_l)-\omega_j
t-\phi_j\,]\}+H.c.\right\}.
\end{eqnarray}
\end{widetext}
Here, $\nu$ and $\nu_l\,(l=1,...,N-1)$ are the frequencies of the
collective center-of-mass (CM) vibrational motion ($l=0$) and the
higher normal modes ($l\geq 1$) of the trapped ions, respectively;
$\hat{a}^\dagger$ and $\hat{a}$ are the ladder operators of CM
mode, while $\hat{b}_l^\dagger$ and $\hat{b}_l$ are the ladder
operators of the higher normal modes.
$\Omega_j\,\,(j=1,2)$ is the carrier Rabi frequency, which
describes the coupling strength between the laser and the $j$th
ion and is proportional to the strength of the applied laser.
$\hat{\sigma}_{z,j}$ and $\hat{\sigma}_{\pm,j}$ are Pauli
operators, $\hbar\omega_0$ is the energy separation of the two
internal states $|g\rangle$ and $|e\rangle$ of the ion, and
$\phi_j$ is the initial phase of the applied laser beam.
The LD parameters $\eta_j$ and $\eta_{j,\,l}$ account for the
coupling strength between the internal state of the $j$th ion and
the vibrational states of the CM mode and the higher frequency
modes, respectively.
Expanding Eq. (1) in terms of creation and annihilation operators
of the normal modes, we can rewrite the Hamiltonian of system in
the interaction picture as
%
\begin{widetext}
\begin{eqnarray}
\hat{H}=\frac{\hbar}{2}\hspace{-0.1cm}\sum_{j=1,2}\hspace{-0.1cm}\left\{\Omega_{j}
\hat{\sigma}_{+,j}\hat{G}_j\left[\exp\left(-\frac{\eta_j^2}{2}-i\phi_j\right)\hspace{-0.2cm}
\sum_{m,n=0}^\infty\hspace{-0.2cm}\frac{(i\eta_j)^{m+n}\hat{a}^{\dagger
m}\,\hat{a}^n}{m!\,n!}\exp[i(m-n)\nu
t+i\delta_jt]\right]\hspace{-0.2cm}+H.c.\right\},
\end{eqnarray}
\end{widetext}
with
\begin{widetext}
$$
\hat{G}_j=\prod_{l=1}^{N-1}\exp\left(-\frac{(\eta_j^l)^2}{2}\right)
\sum_{m',n'=0}^\infty\frac{(i\eta^l_j)^{m'+n'}\hat{b}_l^{\dagger
m'}\,\hat{b}_l^{n'}}{m'!\,n'!}\exp[i(m'-n')\nu_lt].
$$
\end{widetext}
We assume that the frequencies of the applied lasers to be tuned
resonantly on the same lower red-sidebands of the center-of-mass
(CM) vibrational mode, i.e., the frequencies $\omega_j$ of the
applied lasers are chosen to be
$\omega_j=\omega_0-k_j\nu,\,k_j=k=1,2,...$.
Then, like the procedure described in
\cite{Li99,Solano00,Buzek02}, we make the usual rotating wave
approximation (RWA) and have the following effective Hamiltonian
%
\begin{widetext}
\begin{equation}
\hat{H}_{eff}=\frac{\hbar }{2}\sum_{j=1,2} \, \left\{ \Omega
_{j}\hat{F}_j\, \hat{\sigma}_{+,j} \exp\left(\,-
\frac{\eta_j^2}{2}-i\phi_j\right)
\,\,\sum_{n=0}^\infty\frac{(i\eta_j)^{2n+k} \, \hat{a}^{\dagger n}
\, \hat{a}^{n+k}}{n!\,(n+k)!} \, + \, H.c.\,\,\right\},
\end{equation}
\end{widetext}
for small $k$ values. The operator function $$
\hat{F}_j=\prod_{l=1}^{N-1} \exp[-(\eta_j^l)^2/2]
\sum_{n=0}^\infty\frac{(i\eta_j^l)^{2n}}{(n!)^2}\hat{b}_l^{\dagger
n} \hat{b}_l^{n},$$ involving the number operators related to the
higher-$l$ normal modes ($l\geq 1$), is irrelevant because we are
considering the weak excitation regime ($\Omega_j\ll\nu$, i.e.,
the intensity of the applied laser beams is assumed to be
sufficiently weak).
Thus, hereafter we neglect all off-resonant transitions and the
excitations of the higher vibrational modes
\cite{Li99,Solano00,Buzek02}, i.e., let $\hat{F}_j=\hat{I}$, and
only label the CM mode excitations.
We stress the following important fact: the effective Hamiltonian
(3) reduces to that in previous works (e.g.,
\cite{King98,Sackett00,Jonathan01,MS99} under the usual LD
approximation: $(m+1)\eta_j^2\ll 1$), to the lowest order of the
LD parameter $\eta_j$.
Here $m$ is the occupation number of the Fock state of the CM
vibrational quanta.
Also, the Hamiltonian (3) reduces to that in \cite{Li99} for
$k=1$.

In order to manipulate a pair of trapped ions outside the LD
regime, we now wish to solve the quantum dynamical problem
associated with the above Hamiltonian (3) {\it without\/} using
the LD approximation.
All operations presented below are based on this solution and do
not involve quantum transitions to auxiliary atomic levels.
Without loss of generality, the information bus (i.e., the CM
vibrational mode of the ions) is assumed to be prepared beforehand
in a pure quantum state, e.g., the Fock state $|m\rangle$ with
$m<k$.
During the time-evolution
$\hat{U}(t)=\exp(-it\hat{H}_{eff}/\hbar)$, the initial state
$|m\rangle|g_1\rangle|g_2\rangle$ is unchange, i.e.,
\begin{equation}
|m\rangle |g_{1}\rangle |g_{2}\rangle \xrightarrow{\hat{U}(t)}
|m\rangle|g_{1}\rangle|g_{2}\rangle,
\end{equation}
since $\hat{H}_{eff}|m\rangle|g_{1}\rangle |g_{2}\rangle =0$.
Using the relations
$
\hat{H}_{eff}|m\rangle|e_{1}\rangle |g_{2}\rangle
=(-i)^k\hbar\,e^{i\phi_1}\alpha_1|m+k\rangle|g_1\rangle|g_2\rangle,\,\,\,
\hat{H}_{eff}|m\rangle|g_{1}\rangle |e_{2}\rangle
=(-i)^k\hbar\,e^{i\phi_2}\alpha_2|m+k\rangle|g_1\rangle|g_2\rangle,
$
and
$
\hat{H}|m+k\rangle|g_{1}\rangle |g_{2}\rangle
=i^k\hbar\,(e^{-i\phi_1}\alpha_1|m\rangle|e_1\rangle|g_2\rangle+
e^{-i\phi_2}\alpha_2|m\rangle|g_1\rangle|e_2\rangle),
$
we have the evolutions
\begin{widetext}
\begin{eqnarray}
\left\{
\begin{array}{llll}
|m\rangle |g_{1}\rangle |e_{2}\rangle &\xrightarrow{\hat{U}(t)}
&B_{1}(t)|m+k\rangle |g_{1}\rangle |g_{2}\rangle
+B_{2}(t)|m\rangle |g_{1}\rangle |e_{2}\rangle
+B_{3}(t)|m\rangle |e_{1}\rangle |g_{2}\rangle ,\\
\\
|m\rangle |e_{1}\rangle |g_{2}\rangle &\xrightarrow{\hat{U}(t)}
&C_{1}(t)|m+k\rangle |g_{1}\rangle |g_{2}\rangle
+C_{2}(t)|m\rangle |g_{1}\rangle |e_{2}\rangle +C_{3}(t)|m\rangle
|e_{1}\rangle |g_{2}\rangle,
\end{array}
\right.
\end{eqnarray}
\end{widetext}
with $ B_1(t)=(-i)^{k+1}e^{i\phi_2}\frac{\alpha_2\sin(\chi
t)}{\chi}, \,\, B_2(t)=\frac{\alpha_1^2+\alpha_2^2\cos(\chi
t)}{\chi^2}, \,\,
B_3(t)=e^{-i(\phi_1-\phi_2)}\frac{\alpha_1\alpha_2[\cos(\chi
t)-1]} {\chi^2};
C_1(t)=(-i)^{k+1}e^{i\phi_1}\frac{\alpha_1\sin(\chi t)}{\chi},
\,\, C_3(t)=\frac{\alpha_2^2+\alpha_1^2\cos(\chi t)}{\chi^2}, \,\,
C_2(t)=e^{i(\phi_1-\phi_2)}\frac{\alpha_1\alpha_2[\cos(\chi
t)-1]}{\chi^2}. $

Here,
$ \chi=\sqrt{\alpha_1^2+\alpha_2^2},\,\,\,\,
\alpha_j=\Omega_{m,k}^j,\,\, \gamma_j=\Omega_{m+k,k}^{j},\,\,\,
\Omega_{m,k}^j=\frac{\Omega_je^{-\eta_j^2/2}}{2}\sqrt{\frac{(m+k)!}{m!}}\sum_{n=0}^m\frac{(-i\eta_j)^{2n+k}}{(n+k)!}
\left(
\begin{array}{c}
n\\
m \end{array} \right),\,\,\,j=1,2.
$
Finally, in order to obtain the time-evolution of the initial
state $|m\rangle|e_1\rangle|e_2\rangle$, we solve the
Schr\"odinger equation $i\hbar\partial|\psi(t)\rangle/\partial
t=\hat{H}_{eff}|\psi(t)\rangle$ in the invariant subspace
$\{|m+2k\rangle|g_1\rangle|g_2\rangle,|m+k\rangle|e_1\rangle|g_2\rangle,
|m+k\rangle|g_1\rangle|e_2\rangle,|m\rangle|e_1\rangle|e_2\rangle\}$,
\begin{widetext}
$$
i\frac{\partial}{\partial t}\left(
\begin{array}{l}
D_1(t)\\
D_2(t)\\
D_3(t)\\
D_4(t)
\end{array}
\right)=\left(
\begin{array}{llll}
0&(-i)^k\alpha_1e^{i\phi_1}&(-i)^k\alpha_2e^{i\phi_2}&0\\
i^k\alpha_1e^{-i\phi_1}&0&0&(-i)^k\beta_2e^{i\phi_2}\\
i^k\alpha_2e^{-i\phi_2}&0&0&(-i)^k\beta_1e^{i\phi_1}\\
0&i^k\beta_2e^{-i\phi_2}&i^k\beta_1e^{-i\phi_1}&0
\end{array}
\right)\left(
\begin{array}{l}
D_1(0)\\
D_2(0)\\
D_3(0)\\
D_4(0)
\end{array}
\right),
$$
\end{widetext}
with $D_1(0)=D_2(0)=D_3(0)=0,\,D_4(0)=1$. Here,
$|\psi(t)\rangle=D_1(t)|m+2k\rangle|g_1\rangle|g_2\rangle+
D_2(t)|m+k\rangle|e_1\rangle|g_2\rangle+D_3(t)|m+k\rangle|g_1\rangle|e_2\rangle
+D_4(t)|m\rangle|e_1\rangle|e_2\rangle$, and $
\beta_j=\Omega_{m+k,k}^{j}. $ After a long but direct algebraic
derivation, we obtain the exact time-evolution
\begin{widetext}
\begin{eqnarray}
|m\rangle |e_{1}\rangle |e_{2}\rangle &\xrightarrow{\hat{U}(t)}
&D_{1}(t)|m+2k\rangle |g_{1}\rangle |g_{2}\rangle
+D_{2}(t)|m+k\rangle |e_{1}\rangle |g_{2}\rangle
+D_{3}(t)|m+k\rangle |g_{1}\rangle |e_{2}\rangle
+D_{4}(t)|m\rangle |e_{1}\rangle |e_{2}\rangle,
\end{eqnarray}
\end{widetext}
with
$
D_1(t)=(i)^{2k}e^{i(\phi_1+\phi_2)}\frac{\rho}{\Delta}\left[\cos(\lambda_{+}t)-\cos(\lambda_{-}t)\right]$,
\, $D_4(t)=\frac{1}{\Delta}\left
[\zeta_{+}\cos(\lambda_{-}t)\,-\,\zeta_{-}\cos(\lambda_{+}t)\right],
D_2(t)=(-i)^{k+1}e^{i\phi_2}\frac{\rho}{\Delta} \left
[\frac{\alpha_2\rho+\gamma_1\zeta_{+}}{\lambda_{+}\zeta_{+}}\sin(\lambda_{+}t)
-\frac{\alpha_2\rho+\gamma_1\zeta_{-}}{\lambda_{-}\zeta_{-}}\sin(\lambda_{-}t)\right]$,
$ D_3(t)=(-i)^{k+1}e^{i\phi_1}\frac{\rho}{\Delta}
\frac{\alpha_1\rho+\gamma_2\zeta_{+}}{\lambda_{+}\zeta_{+}}\sin(\lambda_{+}t)
-(-i)^{k+1}e^{i\phi_1}\frac{\rho}{\Delta}\frac{\alpha_1\rho+\gamma_2\zeta_{-}}{\lambda_{-}\zeta_{-}}\sin(\lambda_{-}t).
$
Here,
$
\rho=\alpha_1\beta_2+\alpha_2\beta_1,\,\,\,\,
\zeta_{\pm}\,=\,\lambda_{\pm}^2-\sum_{j=1}^2\alpha_j^2,\,\,\,\,
\lambda_{\pm}\,=\,\sqrt{\frac{\Lambda\pm\Delta}{2}},\,\,\,\
\Lambda\,=\,\sum_{j=1}^2(\alpha_j^2+\beta_j^2),\,\,\,\,\,
\Delta^2\,=\,\Lambda^2-4(\alpha_1\beta_1-\alpha_2\beta_2)^2\,\,\,.
$
%
%

Analogously, the effective Hamiltonian and the relevant dynamical
evolutions for other driving cases can also be derived exactly.
For example, for the case where $k_1=k_2=k'<0$ and $|k'|=k>m$
(i.e., the ions are excited by blue-sideband laser beams with
equal frequencies instead of red-sideband ones), the effective
Hamiltonian and the relevant dynamics of the system can be easily
obtained from Eq.~(3) and Eqs.~(4-6) by making the replacements:
$\hat{a}\longleftrightarrow\hat{a}^\dagger$ and
$|e_j\rangle\longleftrightarrow |g_j\rangle$, respectively.

\section{Manipulation of quantum information in two trapped cold ions}

Based on the conditional quantum dynamics for the two-qubit system
derived in the previous section, we now show how to effectively
manipulate two-qubit quantum information stored in two ions by
applying a pair of simultaneous laser pulses.
Generally, the motion state entangles with the spin states during
the dynamical evolution.
We afterwards focus on how to decouple them and realize {\it in
one step} and beyond the LD limit: either a two-qubit controlled
operation or an entanglement between the trapped ions.
The state of the information bus (CM mode) remains in its initial
state, which is not entangled with the qubits after the
operations.
This is achieved by properly setting up the controllable
experimental parameters, e.g., the Lamb-Dicke parameters $\eta_j$,
the efficient Rabi frequencies $\Omega_j$, the frequencies
$\omega_j$ ($j\,=\,1,2$) and the duration of the applied
synchronous pulses.

\subsection{Two-Qubit Controlled Operations}

As one of the simplest universal two-qubit quantum gates, the
$C^Z$ logic operation between the $1$st and the $2$nd ions
\begin{widetext}
\begin{eqnarray}
\hat{C}^Z_{12}\,\,=\,\,|g_{1}\rangle |g_{2}\rangle \langle
g_{1}|\langle g_{2}|\,+\,|g_{1}\rangle |e_{2}\rangle \langle
g_{1}|\langle e_{2}|\,+\,|e_{1}\rangle |g_{2}\rangle \langle
e_{1}|\langle g_{2}|\,-\,|e_{1}\rangle |e_{2}\rangle \langle
e_{1}|\langle e_{2}|,
\end{eqnarray}
\end{widetext}
means that if the first ion is in the state $|g_1\rangle $, the
operation has no effect, whereas if the control qubit (first ion)
is in the state $|e_1\rangle $, the state of the second ion is
rotated by the Pauli operator $\hat{\sigma}_z$.
The first qubit is the control qubit and the second one is the
target qubit. The  $C^Z$ operation is also known as
controlled-rotation (CROT).
It is seen from Eqs.~(4-6) that the $\hat{C}_{12}^Z$ gate can be
implemented exactly by a {\it one-step} operation, if the
experimental parameters are set up so that the following
conditions are simultaneously satisfied,
\begin{equation}
\cos(\chi\tau_z)=1,\,\,\,\,\,\,\,\cos(\lambda_{+}\tau_z)=\cos(\lambda_{-}\tau_z)=-1.
\end{equation}
Here $\tau_z$ is the duration of the two applied synchronous
pulses. Notice that the second condition in (8), on
$\lambda_{\pm}$, is equivalent to requiring $|D_4(t)|^2=1$, which
forces $|D_1(t)|^2=|D_2(t)|^2=|D_3(t)|^2=0$ due to normalization.
The information bus remains in its initial state after the
operation. Without loss of generality, we give some solutions of
the conditional equations (8) for $m=0$ in Table I.

We see from the Table I that both the small and large, both the
negative and positive, values of the LD parameters may be chosen
to satisfy the conditions (8) for realizing the desired two-qubit
controlled gate.
Our approach does not assume the LD approximation where $\eta_j\ll
1$ for $m=0$.
Thus, the present scheme can operate outside the LD regime and
$\eta_j$ can be large.

According to previous works (see, e.g., \cite{CZ95,wei02}), it is
known that an exact two-qubit gate $\hat{C}^Z$ surrounded by two
one-qubit rotations on the target qubit can give rise to an exact
two-qubit CNOT gate $\hat{C}^X$.
The present work shows that the CNOT gate between different ions
can be realized by using a {\it three-step} pulse process, which
is simpler than the previous schemes which usually employed either
{\it five} pulses \cite{CZ95,wei02}, {\it six} pulses
\cite{Jonathan00} or {\it seven\/}-step operations
\cite{Childs01}.
\begin{widetext}
\begin{center}
\begin{tabular}{|c|c|c|c|c|c|c|c|c|}
\hline $\Omega_2/\Omega_1$ & $\,\,k\,\,$ & $\eta_1=\eta_2=\eta$ &
$\Omega_1 \tau_z$& $\Omega_2/\Omega_1$ & $\,\,k\,\,$ &
$\eta_1=\eta_2=\eta$ &
$\Omega_1 \tau_z$\\
\hline $  2.03951    $&  $  1    $ & $ 1.93185 $ &$ 18.5069 $&
$ 1.81182     $ & $  1    $ & $  2.30578 $& $ 37.5859 $\\
\cline{3-4}\cline{6-8} $                 $ & $        $& $0.517638
$& $12.2197 $&
$                 $&  $   2    $& $ \pm
0.253727$ & $137.757$\\
\cline{2-4}\cline{7-8}
 $                 $&  $   2   $ & $ \pm 0.915272 $
&$  14.1979      $&
$                 $&$         $ & $ \pm
2.81702     $&  $ 57.2117$\\
\cline{3-4}\cline{6-8} $                 $ & $       $&  $\pm
2.67624 $&$39.2315$ &
$                 $& $   3     $ & $
0.859544   $& $33.8887$\\
\cline{2-4}\cline{7-8} $                 $&  $   3   $ & $ 1.12532
$ &$17.9115$ &
$                 $&  $       $ & $ 3.40669
$& $124.419$\\
\cline{3-4}\cline{5-8} $                 $ & $       $& $2.69702
$&$26.2324      $ &
$     4.02791 $& $   1    $ & $
1.87083               $& $ 18.6274      $\\
\cline{3-4}\cline{7-8} $                 $ & $       $& $3.34152
$& $96.5506     $&
$                 $ &$       $ & $  0.707107              $ & $ 10.9967     $\\
\cline{1-4}\cline{6-8} $ 0.658331     $ &$    1   $ & $ 1.76579 $
& $56.5182       $                   &
$ $&$  2    $ & $  \pm
0.983608 $ & $ 14.3592  $\\
\cline{3-4}\cline{7-8}
 $ $ &$        $ & $  0.939131 $& $34.7414
$  &
$ $& $   $&$  \pm 2.65189     $&  $ 40.9890    $\\
\cline{2-4}\cline{6-8} $                 $&$    2    $ & $ \pm
1.09276           $&  $    45.1673$ $                $&
$ $& $ 3    $ & $ 1.16543               $ &$ 18.4811 $
\\
\cline{3-4}\cline{7-8} $                 $&$         $ & $ \pm
2.60881           $&  $    131.088$ &
$                 $ & $ $&  $2.63899            $& $26.2548 $
\\
\cline{2-4}\cline{7-8} $                 $&$    3    $ & $ 1.23348
$&  $    58.6299$ &
$                 $ &$ $ & $  3.11088 $ & $62.2365 $
\\
\cline{3-4}\cline{7-8} $                 $&$         $ & $ 2.55336
$&  $    80.4486$&
$                 $ &$       $ & $ 3.33069              $&
$102.936    $
\\
\cline{1-8} \hline
\end{tabular}
\end{center}
\begin{center}
TABLE I. A few solutions of equations (8) for $m=0$. These
parameters realize a $C^Z$ or CROT gate between two trapped ions.
Here, $\tau_z$ is the duration of the applied pulses, $\Omega_j$
and $\eta_j$ ($j=1,2$) are the effective Rabi frequencies and the
Lamb-Dicke parameters, respectively. The frequencies of the
applied pulses are equal, i.e., $k_1=k_2=k=1,2,3$.
\end{center}
\end{widetext}

\subsection{Two-Qubit Entangled States}

Recently, the quantum entanglement of two and four trapped ions
have been generated experimentally (see, e.g.,
\cite{Turchette98,Sackett00}), although the operations are limited
to the weak-coupling LD regime. We now show that the entangled
states of two trapped ions can also be produced outside the LD
limit.
Indeed, the dynamical evolutions (4-6) clearly reveal that there
are many ways to produce various deterministic entangled states of
two trapped ions.
For example, if the conditions
\begin{equation}
\alpha_j=\beta_j, \,\,\,\,\phi_1=\phi_2,\,\,\,\,\,
\alpha_1\neq\alpha_2,\,\,\,\,\,\,\,\cos(\chi\tau_e)=-1,
\end{equation}
are satisfied, then two equal red-sideband pulses (i.e.,
$k_1=k_2=k\,>\,0$) with frequencies
$\omega_1\,=\,\omega_2\,=\,\omega_0\,+\,k\,\nu$, applied to two
ions individually and simultaneously, yield the following
dynamical evolutions:
\begin{widetext}
\begin{eqnarray}
\left\{
\begin{array}{ll}
\vspace{0.4cm} &|m\rangle\,|g_1\rangle|e_2\rangle\,\longrightarrow
\,|m\rangle\otimes|\psi_{12}^- \rangle,\,
|\psi^{-}_{12}\rangle=U\,|g_1\rangle|e_2\rangle
\,-\,V\,|e_1\rangle|g_2\rangle,\\
&|m\rangle\,|e_1\rangle|g_2\rangle\,\longrightarrow
\,|m\rangle\otimes|\psi^{+}_{12}\rangle,\,
|\psi^{+}_{12}\rangle=-V\,|g_1\rangle|e_2\rangle
\,-\,U\,|e_1\rangle|g_2\rangle,
\end{array}
\right.
\end{eqnarray}
\end{widetext}
with
$$
U=\frac{\alpha_1^2-\alpha_2^2}{\alpha_1^2+\alpha_2^2},\,\,\,\,\,\,\,\,\,
V=\frac{2\alpha_1\,\alpha_2 }{\alpha_1^2+\alpha_2^2}.
$$
Therefore, entangled states $|\psi^{\pm}_{12}\rangle$ can be
generated, in a {\it single-step} process, by the dynamical
evolution of the initial non-entangled states
$|g_1\rangle\,|e_2\rangle$ or $ |e_1\rangle\,|g_2\rangle$. We note
that the degrees of entanglement for the above entangled states
$|\psi^{\pm}_{12}\rangle$ are equivalent. They are
\begin{equation}
E \, = \, -U^2 \,log_2\,U^2- V^2 \,log_2\,V^2 .
\end{equation}
Here $E$ is the degree of entanglement defined \cite{Bennett96} as
$
E[\psi] \, = \, - \sum_i\,C_i^2\,\,log_2C_i^2,
$
for an general two-particle entangled pure state
$
|\psi(A,B)\rangle \; = \;
\sum_i\;C_i\;|\alpha_i\rangle_A\,\otimes\,|\beta_i\rangle_B,\,\,\,\,\,\,\,\sum_i\,|C_i|^2\,=\,1.
$
The entangled state of two trapped ions realized experimentally in
\cite{Turchette98} is not the maximally entangled state. Its
degree of entanglement $E$ is $0.94$. However, in principle,
maximally entangled states with $E=1$ can be generated
deterministically in the present scheme. Indeed, it is seen from
figure 1 that, if the experimental parameters further satisfy the
following conditions
\begin{equation}
\frac{\alpha_2}{\alpha_1}=\sqrt{2}\pm\,1 ,\,\,\,\,\,\,\,\,
\alpha_1\tau_e=\frac{(2l-1)\pi}{\sqrt{4\pm\,
2\sqrt{2}}},\,\,\,l=1,2,3...,
\end{equation}
the above entangled states $|\psi^{\pm}_{12}\rangle$ become the
maximally entangled two-qubit (i.e., EPR) states
\begin{eqnarray}
|\Psi^{\pm}_{12}\rangle=\frac{1}{\sqrt{2}}\,\left(|g_1\rangle|e_2\rangle\pm\,|e_1\rangle|g_2\rangle\right).
\end{eqnarray}

In Fig.~1, we plot the degree of entanglement $E$ for the above
entangled states versus the ratio $\alpha_2/\alpha_1$, which is a
function of the LD parameters $\eta_j$, efficient Rabi frequencies
$\Omega_j$, and the laser frequencies $\omega_j$, ($j=1,2$).
\begin{figure}
\vspace{2.2cm}
\includegraphics[height=6cm,width=12cm] {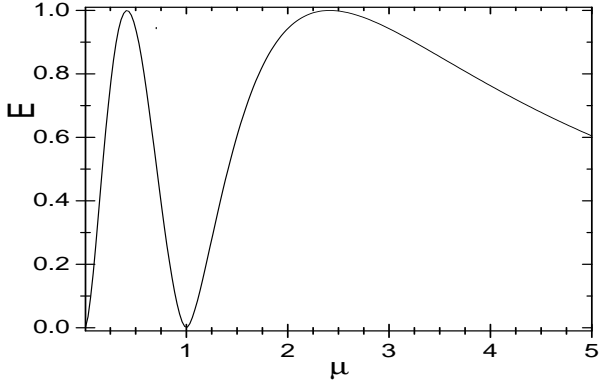}
\vspace{-3cm} \caption{\label{fig1}The degree of entanglement $E$
for the states $|\psi^{\pm}_{12}\rangle$ versus
$\mu\,=\,\alpha_2/\alpha_1$, which is a function of the LD
parameters $\eta_j$, effective Rabi frequencies $\Omega_j$, and
the laser frequencies $\omega_j$, ($j=1,2$). Note that $E\,=\,1$
for $\alpha_2/\alpha_1\,=\,\sqrt{2}\,\pm\,1$, and $E\,=\,0$ for
$\alpha_2/\alpha_1\,=\,1$. For
$\alpha_2/\alpha_1\,>\,\sqrt{2}\,+\,1$ the degree of entanglement
$E$ decreases when $\alpha_2$/$\alpha_1$ increases.}
\end{figure}
Explicitly,
\begin{widetext}
\begin{equation}
\frac{\alpha_2}{\alpha_1}=
\frac{\Omega_2}{\Omega_1}\,\left(\,\frac{\eta_2}{\eta_1}\,\right)^{|k|}\exp\left(-\frac{\eta_2^2-\eta_1^2}{2}\right)
\,\times\,\frac{\sum_{n=0}^m(-i\eta_2)^{2n}C_m^n/(n+|k|)!}
{\sum_{n=0}^m(-i\eta_1)^{2n}C_m^n/(n+|k|)!},
\end{equation}
\end{widetext}
for the processes (10). Especially, for the commonly considered
case, $m\,=\,0$, the above equation becomes
\begin{equation}
\frac{\alpha_2}{\alpha_1}\,=\,
\frac{\Omega_2}{\Omega_1}\,\left(\,\frac{\eta_2}{\eta_1}\,\right)^{|k|}\exp\left(-\,
\frac{\eta_2^2-\eta_1^2}{2}\right).
\end{equation}
Obviously, the values of $E$ depend on the choice of the
experimental parameters $\Omega_j$, $\eta_j$ and $k$ (thus
$\omega_j$), ($j=1,2$). For $\alpha_2/\alpha_2\,=\,1$ the
condition (9) is violated and thus $E\,=\,0$. Similarly,
\begin{equation}
\lim_{\alpha_2/\alpha_1\rightarrow \infty}U=1,\,
\lim_{\alpha_2/\alpha_1\rightarrow \infty}V=0,\,\,{\rm
thus,}\,\lim_{\alpha_2/\alpha_1\rightarrow \infty}\,E\,=\,0.
\end{equation}
Inversely, it is seen from Eqs.~(12-15) that $E\,=\,1$ for
$\alpha_2/\alpha_1\,=\,\sqrt{2}\pm 1$. This implies that two-qubit
maximally entangled states can be generated deterministically by
using a {\it single-step} operation beyond the LD limit. For
example, if the experimental parameters are set up simply as $
\eta_1=\eta_2,\,\,\Omega_2/\Omega_1=\sqrt{2}\pm 1, $ the EPR state
$|\Psi_{12}^{-}\rangle$ can be generated by using a {\it
single-step\/} synchronous red-sideband $\pi$ pulses with
frequencies $\omega_1=\omega_2=\omega_0+\nu$ and duration $
\tau_e=\pi/(\alpha_1\sqrt{4\pm 2\sqrt{2}}).$
%
%

\section{Conclusions and discussions}

Based on the conditional quantum dynamics for two-qubit system, we
have shown that, under certain conditions, the quantum controlled
gate or entanglement between a pair of trapped ions can be
realized deterministically by {\it only a single-step operation},
performed by simultaneously applying two laser pulses to two ions.
Each of the laser beams interacts with a single ion.
Neither auxiliary atomic level nor Lamb-Dicke approximation are
required during the operation.
The CM mode of the ions always remains in its initial quantum
state after the operation.

We now give a brief discussion on the experimental realization of
the present scheme.
For ion-trap quantum information processing, the information bus,
i.e. the usual collective CM vibrational mode, must first be
initialized in a pure quantum state, e.g., its ground state.
Recently, the collective motion of two and four $^9$Be$^+$ ions
has been successfully cooled to its ground state
\cite{King98,Sackett00}.
This is a further step towards realizing the ion trap quantum
computer.

Although all relevant experiments are operated in the LD regime,
we show that the coherent manipulation of two trapped cold ions
may also be implemented outside the LD limit. According to the
experiment reported in \cite{Turchette98}, two $^9$Be$^+$ ions
spaced $2\mu$m apart can be confined along the axis of a linear
Paul trap with an axial center-of mass frequency of
$\nu\,\,=\,\,\omega_z\,\,\approx\,
7\,\,$\,MHz\,\,$\ll\,\,\omega_{x,\,y}$.
The collective mode of motion of the ions has been successfully
cooled to the ground state by Doppler cooling\,(see, e.g.,
\cite{Wineland79}\,).
The qubit states are
$2s^2S_{1/2}\,|F=2,\,m_F=2\rangle\,\,=\,\,|g\rangle$ and
$2s^2S_{1/2}|F=1,m_F=1\rangle\,\,=\,\,|e\rangle$ separated in
frequency by $\omega_0/2\pi\approx 1.25$\,\,GHz
\cite{Turchette98}.
According to the above analysis, we note that the present
theoretical scheme may be realized by current experimental
technologies by properly setting up the controllable parameters
(\,e.g. $\eta_j$,\,\, $\Omega_j$,\,\, $\omega_j$\,) and the
durations of the applied laser pulses. Indeed, it is seen from the
formulae (see, e.g., \cite{Steane98,Buzek02})
\begin{equation}
\eta_j\,=\,\cos\theta_j\,\sqrt{\frac{\hbar
\kappa_j^2}{2MN\nu}},\,\,\,\,\,\,
\theta_j\,=\,\arccos\left(\frac{\vec{\kappa_j}\cdot\vec{z}_j}{\kappa_j}\right).
\end{equation}
that the LD parameter $\eta_j$ can be controlled by adjusting the
wave vector $\vec{\kappa}_j$ of the applied laser pulse.
Obviously, $\eta_j$ can be positive or negative, depending on the
values of $\theta_j$. Here, $MN$ is the total mass of the ion
chain and $\theta_j$ is the angle between the laser beam and the
$z$ axis. The efficient Rabi frequency \cite{Turchette98} of the
$j$th ion,
\begin{equation}
\Omega_j\,=\,\Omega_c\,J_0\,\left(|\vec{\delta
\kappa_j}|\,\xi_j\right),
\end{equation}
can be chosen properly for separate addressing and satisfying the
parameter conditions (8), (9), and (12) for different operations.
Here $J_0$ is the zero-order Bessel function and $\xi_j$ is the
amplitude of the motion along $x$ associated with ion $j$
\cite{Turchette98}.
Jefferts {\ et al.} \cite{Jeff95} point out that applying a static
electric field to push the ions along $x$ may control the
amplitude of $\xi_j$ and thus the $\Omega_j$ of the ion
micro-motion.

Finally, we note that the duration of the two applied simultaneous
pulses for realizing the above quantum controlled operation is not
much longer than that for other schemes (see, e.g.,
\cite{King98,Sackett00,Jonathan01,MS99}) operating in the LD
regime.
The shortest duration of the applied synchronous pulses for
realizing the above manipulations of two trapped ions is about
$10^{-4}$ seconds,
of the same order of the gate speed operating in the LD regime
\cite{Monroe95},
for $\Omega_1/2\pi\,\approx\,\,225\,k$Hz \cite{Turchette98}.
Of course, to excite only the chosen sidebands of the CM mode, the
spectral width of the applied laser pulse has to be sufficiently
small.
It might seem at first, from the above numerical results, that the
present scheme for realizing the desired gate operation cannot be
easily implemented, as the relevant experimental parameters should
be set up accurately.
However, this is not the case.
Simple numerical analysis shows that the lowest probability of
realizing the desired operation is still very high, even if the
relevant parameters cannot be set up exactly.
For example, the lowest probability of realizing the two-qubit
$\hat{C}^Z$ operation is up to $99.97\%$ ($99.49\%$), if the rate
of the two efficient Rabi frequencies $\Omega_2/\Omega_1$ is
roughly set up as $2.03$ ($2.0$), which is $0.5\%$ ($1.9\%$) away
from the exact solution of condition (8), see table I.
Therefore, based on the current ion-trap technologies the proposed
scheme may be realizable in the near future.

\section*{Acknowledgments}
We acknowledge X. Hu for useful discussions, C. Monroe for
comments on the manuscript, and the partial support of ARDA,
AFOSR, and the US National Science Foundation grant No.
EIA-0130383.


\begin{references}
\bibitem{Bennett00}
M.A.~Nielsen, I.L.~Chuang, \textit{Quantum Computation and Quantum
Information}, (Cambridge University Press, Cambridge, September
2000); C.H. Bennett and D.P. DiVincenzo, Nature, {\bf 44} (2000)
247; A. Ekert and R. Josza, Rev. Mod. Phys. {\bf 68} (1996) 733 ;
A.M. Steane, Rep. Prog. Phys. {\bf 61} (1998) 117 .
%
\bibitem{Bell87}
A. Einstein, B. Podolsky, and N. Rosen, Phys. Rev. {\bf 47} (1935)
777; J. S. Bell, {\ Speakable and Unspeakable in Quantum
Mechanics}\,(Cambridge University Press, Cambridge, England,
1987); M. A. Rowe {\it et al}, Nature, {\bf 409} (2001) 791.
%
\bibitem{Shor94}
P. Shor, in Proceedings of the 35th Annual Symposium on the
Foundations of Computer Science, edited by Shafi Goldwasser (IEEE
Computer Society Press, New York, 1994), p. 124.
%
\bibitem{Kwiat95}
P.G. Kwiat {\it et al.}, Phys. Rev. Lett. {\bf 75} (1995) 4337 ;
D. Bouwmeester {\it et al.}, Phys. Rev. Lett. {\bf 82} (1995)
1345.
%
\bibitem{Sackett00}
H. C. N\"agerl {\it et al.}, Phys. Rev. A {\bf 60} (1999) 145 ; C.
A. Sackett {\it et al.}, Nature (London) {\bf 404} (2000) 256.
%
%
\bibitem{CZ95}
J.I. Cirac and P. Zoller, Phys. Rev. Lett. {\bf 74} (1995) 4091.
%
\bibitem{KS01}
D. Kielpinski {\it et al.}, Science {\bf 291} (2001) 1013.
%
\bibitem{Monroe95}
C. Monroe {\it et al.}, Phys. Rev. Lett. {\bf 75} (1995) 4714; Ch.
Roos {\it et al.}, Phys. Rev. Lett. {\bf 83} (1999) 4713.
%
\bibitem{King98}
B.E. King {\it et al.}, Phys. Rev. Lett. {\bf 81} (1998) 1525;
%
\bibitem{Turchette98}
Q.A. Turchette {\it et al.}, Phys. Rev. Lett. {\bf 81} (1998)
3631.
%
\bibitem{Duan01}
L.M. Duan, J.I. Cirac, and P. Zoller, Science {\bf 292} (2001)
1695; S.B. Zheng, Phys. Rev. A{\bf 65} (2002) 051804.
%
\bibitem{Jonathan01}
D. Jonathan and M.B. Plenio, Phys. Rev. Lett. {\bf 87} (2001)
127901.
%
\bibitem{wei02}
L.F. Wei, S.Y. Liu and X.L. Lei, Phys. Rev. A {\bf 65} (2002)
062316; Opt. Commun. {\bf 208} (2002) 131.
%
\bibitem{sharma03}
S.S. Sharma and A. Vidiella-Barranco, Phys. Lett. A {\bf 309}
(2003) 345.
%
\bibitem{Jonathan00}
D. Jonathan, M.B. Plenio, and P.L. Knight, Phys. Rev. A {\bf 62}
(2000) 042307.
%
\bibitem{Childs01}
A.M. Childs and I.L. Chuang, Phys. Rev. A {\bf 63} (2000) 012306.
%
\bibitem{MS99}
K. M\o lmer and A. S\o rensen, Phys. Rev. Lett. {\bf 82} (1999)
1835; A. S\o rensen and K. M\o lmer, {\it ibid}. {\bf 82} (1999)
1971; Phys. Rev. A {\bf 62} (1999) 022311; A. Svandal and J.P.
Hansen, Phys. Rev. A {\bf 65} (2002) 033406.
%
\bibitem{Meekhof96}
D.M. Meekhof {\it et al.}, Phys. Rev. Lett. {\bf 76} (1996) 1796;
D. Leibfried {\it et al.}, {\it ibid}, {\bf 77} (1996) 4281; {\bf
89} (2002) 247901; J.I. Cirac {\it et al.}, {\it ibid}. {\bf 70}
(1993) 762; {\bf 70} (1993) 556.
%
\bibitem{Wineland79}
D.J. Wineland and W.M. Itano, Phys. Rev. A {\bf 20} (1979) 1521.
%
\bibitem{Steane98}
D. Stevens, J. Brochard and A.M. Steane, Phys. Rev. A{\bf 58}
(1998) 2750; A. Steane, App. Phys. {\bf B 64} (1997) 623; D. F. V.
James, $ibid$ {\bf 66} (1998) 181.
%
\bibitem{Morigi99}
G. Morigi, J. Eschner, J.I. Cirac, and P. Zoller, Phys. Rev. A
{\bf 59} (1999) 3797.
%
%
%
\bibitem{leibfried99}
D. Leibfried, Phys. Rev. A {\bf 60} (1999) R3335.
%
\bibitem{Solano00}
E. Solano {\it et al}, Phys. Rev. A{\bf 62} (2000) 021401; {\bf
59} (1999) R2539; {\bf 64} (2001) 024304.
%
\bibitem{Li99}
L.X. Li and G.C. Guo, Phys. Rev. A {\bf 60} (1999) 696.
%
%
\bibitem{Buzek02}
M. \u Sa\u sura and V. Bu\u zek, J. Mod. Opt., {\bf 49} (2002)
1593; H.Ch. N$\ddot{a}$gerl, {\it et al}, Forts. Phys., {\bf 48}
(2000) 623.
%
%
\bibitem{Bennett96}
C.H. Bennett, H.J. Bernstein, S. Popescu, and B. Schumacher ,
Phys. Rev. A {\bf 53} (1996) 2046.
%
\bibitem{Jeff95}
S.R. Jefferts, C. Monroe, E.W. Bell, and D.J. Wineland, Phys. Rev.
A {\bf 51} (1995) 3112.
%
%
%
\end{references}
\end{document}